\documentclass[11pt]{article}
\newcommand{\be}{\begin{equation}}
\newcommand{\ee}{\end{equation}}
\usepackage{amsmath}
\usepackage{amsfonts}
\usepackage{amssymb}

\usepackage{cite}
\usepackage{graphicx}
\begin{document}
	
	\begin{center}
		{\Large{\textbf{Comment on}}}
		
		{\Large{\textbf							
		{\emph{Energy Non-Conservation in Quantum Mechanics},}}}
	
		{\Large{\textbf 
		{\emph{arXiv:2101.11052v2 [quant-ph]} }}}\\ 
	
		\vspace{2mm}
		 Johann Summhammer\footnote[1]{Email: johann.summhammer@tuwien.ac.at} \\ 
		Technische Universit\"at Wien\\
				Institute of Atomic and Subatomic Physics\\ 
				Stadionallee 2, 1020 Vienna, Austria\\

	\end{center}

\begin{abstract}
In the posting arXiv:2101.11052v2 [quant-ph] an experimental protocol involving two spins is proposed, which should show violation of energy conservation in a quantum experiment. In the present comment an unjustified mathematical approximation leading to that conclusion is pointed out. A detailed analysis will restore energy conservation. 

\end{abstract}

The posting cited in the title \cite{CA}, and also published in \cite{Carroll2021}, suggests an experiment with two spins, with the purpose of showing non-conservation of energy in quantum mechanics. I will refer to \cite{CA} and \cite{Carroll2021} as ENC and will argue that there is an error in the evaluation of the Hamiltonian. When doing a correct analysis, energy will be conserved for all possible outcomes of the experiment. I will also mention a conceptually similar experiment, for which the Hamiltonian and the solutions have been worked out in detail. The layout of the experiment suggested in ENC is shown in figure 1.
\begin{figure}[h]
	\centering
	\includegraphics[scale=0.47]{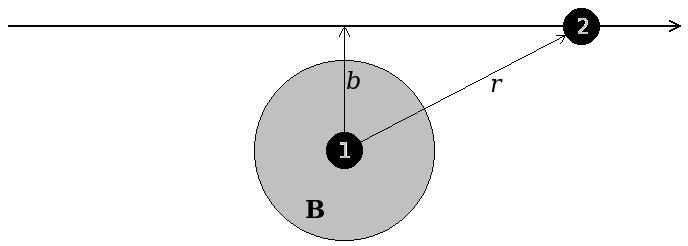}
	\caption{Particle 1 is trapped in the magnetic field $\bf{B}$, while particle 2 is flying by at a distance $b$. Both particles have spin 1/2 and become entangled by the spin-spin interaction. Particle 2 does not enter the region of the magnetic field.}
	\label{}
\end{figure}

\noindent
The core of the proposal of ENC is to let two spin-1/2 particles interact by their magnetic moments. Particle 1 is trapped in a magnetic field parallel to the z-axis, and is in a superposition of its two spin eigenstates in the field, which are also eigenstates of energy. Particle 2 passes near particle 1, but never enters the magnetic field region. It is also in a superposition of the two spin eigenstates relative to the z-axis. While particle 2 is flying by, the two particles will become entangled in spin. Once particle 2 is sufficiently far from the interaction region, a measurement of its spin is performed. The result will be either "up" or "down". Because of the entanglement particle 1 must then be in exactly one of its two possible eigenstates in the magnetic field. But these two states of particle 1 have very different energy, depending on the magnitude of B. In ENC it is argued, that the back-reaction on particle 2 during the interaction is negligible from the energetic point of view and can be ignored. Therefore, the total energy of the system would be different, depending on the measurement result, and this would constitute a violation of energy conservation.

Let us look at the problem in more detail. The Hamiltonian given in ENC (eq.16 in \cite{CA}) is
\be
\hat{H} = -\gamma B_z S_1^z \otimes 1 \hspace{-1.2mm} 1_2 + \frac{g}{r^3} \left[ \vec{S}_1 \cdot \vec{S}_2 -3( \vec{S}_1 \cdot \vec{r})( \vec{S}_2 \cdot \vec{r})  \right].
\ee

\noindent
This Hamiltonian has no explicit time dependence, therefore it commutes with the energy operator $-i\hbar \partial / \partial t$, and total energy will be conserved, no matter which particular initial wavefunction for the combined system of the two spins is assumed. This is commented on in ENC, but seems to be understood as valid only for expectation values rather than all individual outcomes of a measurement. On the other hand, the Hamiltonian does depend on the position $\vec{r}$ of particle 2, and therefore it does not commute with the momentum operator for particle 2, which involves the derivative with respect to $\vec{r}$. As a consequence, particle 2 may be found, in general, in different momentum states, once it is far away from the interaction region. This is observed correctly in ENC, but there it is argued, that the difference in kinetic energy of these states  can be made negligible relative to the energy difference which the two possible eigenstates of particle 1 will have in the magnetic field B, because the magnetic field can be made arbitrarily large. Hence, it is argued in ENC, if one measures the spin of particle 2, one will be able to infer the spin state of particle 1, and since these states will be associated with very different energies, while the different energies of the states of particle 2 are negligible, energy must have appeared or disappeared.

Clearly, in view of the static Hamiltonian this conclusion must be wrong. Aside from this formal argument, one can also give a more physical one relying on classical intuition. Imagine a classical particle 1 suspended in the  magnetic field, and for the sake of simplicity, let its magnetic moment be anti parallel to the magnetic field, so that it has minimum potential energy. Now along comes classical particle 2 with its magnetic moment pointing in some other direction. When it passes near particle 1, the two particles will exert a mutual torque on each other, and this will generally change the final direction of their respective magnetic moments. Classically, there can be no doubt, that the amount, by which particle 2 will be able to change the direction of the magnetic moment of particle 1, \emph{will depend on the magnitude of the magnetic field}. If B is very strong, particle 1 will be so rigidly bound in its state, that particle 2 will have a hard time to cause any change at all. And if it does manage to influence particle 1, any change of potential energy of particle 1 in the magnetic field will be compensated by a change of the final linear momentum of particle 2. These considerations will also fit together nicely with conservation of total angular momentum, because any change of linear momentum of particle 2 implies a change of its orbital angular momentum with respect to the position of particle 1. Based on this classical intuition we expect  something similar to happen in the quantum mechanical case.

The mistake of ENC must consist in neglecting the back reaction of field-trapped particle 1 on the final momentum states of particle 2. But it is not easily recognizable how the magnetic field should become manifest in the final states of particle 2, when this particle remains outside the field region all the time. At least, it is not immediately obvious from the Hamiltonian of eq.1, where $\vec{S}_2$ only interacts with $\vec{S}_1$. I did not find papers which have solved the problem of ENC, but I did come across a paper by P. L. Nash, which analyzes a very similar problem \cite{Nash1996}. In particular, that paper makes visible how the potential, to which one particle is exposed, becomes reflected in another particle, which interacts with it, but which is not exposed to the potential itself. The scheme is sketched in figure 2.
\begin{figure}[h]
	\centering
	\includegraphics[scale=0.50]{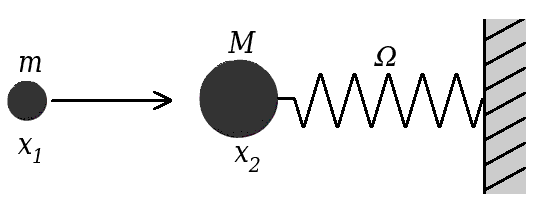}
	\caption{The conceptual scheme considered in \cite{Nash1996}. A particle with mass $m$ is scattered at another particle with mass $M$, which resides in a harmonic oscillator potential.}
	\label{}
\end{figure}

A particle with mass $M$ is trapped in a harmonic oscillator potential characterized by the angular frequency $\Omega$. Another particle with mass $m$ flies toward the trapped particle and is scattered. The problem is analyzed in one dimension. Here, the free particle has index 1, and the particle oscillating in the trap has index 2. The Hamiltonian given by Nash reads (eq.1 in \cite{Nash1996})
\be
H = \frac{1}{2M}p_2^2 + \frac{1}{2}M\Omega^2 x_2^2 + \frac{1}{2m}p_1^2 + V(x_1 - x_2).
\ee 
The interaction potential between the particles is assumed symmetric, $V(x_1 - x_2) = V(x_2 - x_1)$. It is noteworthy, that it only depends on the distance between the two particles, but does not depend on the parameter characterizing the harmonic trap, $\Omega$. This is exactly as in ENC. A further similarity with ENC is the time independence of the Hamiltonian.
In the analysis Nash makes use of the fact that the solutions of the quantum mechanical harmonic oscillator are essentially the Hermite polynomials times an exponential prefactor \cite{WikiHarmonicOsc}. Therefore it may be convenient to introduce a dimensionless position coordinate $X$ for particle 2 defined by
\be
X = x_2 \sqrt{\frac{M\Omega}{\hbar}},
\ee
and expand the potential between the two particles in Hermite polynomials
\be
V\left(x_1 - X \sqrt{\frac{\hbar}{M \Omega}} \right) = \sum_n H_n(X) V_n(x_1).
\ee
The $V_n(x_1$) are expansion coefficients depending on the coordinate of particle 1 and have to be calculated. Here one can already see how the angular frequency $\Omega$ of the harmonic oscillator enters into the potential to which particle 1 will actually be subjected when interacting with particle 2. This is quite surprising, because the potential between the two particles is only assumed as some function of the distance between them. And at the face of it, the angular frequency $\Omega$ of the harmonic oscillator, which only concerns particle 2, should not play a role for particle 1. 

Nash then proceeds to calculate the expansion coefficients $V_n(x_1)$ and from them derives the scattering matrix. From the scattering matrix the probabilties can be determined for the exchange of an integer number of vibrational energy quanta $\hbar \Omega$ between the oscillating mass $M$ and the scattered mass $m$. For a given number $n$ of quanta given off by $M$ the same number of quanta is absorbed by $m$, or vice versa. The question of violation of conservation of energy does not even arise, because the correct solution of the time independent Hamiltonian ensures it for each conceivable final state of the system. 

For the special case of a hard sphere potential Nash gives explicit expressions for the elements of the scattering matrix. These are cumbersome terms. But all of them contain a scaling factor of the form $\sqrt{2M\Omega/\hbar}$. This shows the importance of the frequency of the harmonic oscillator, and thus of the stiffness of the spring constant, for the various transition probabilities. If a similarly rigorous derivation is done for the case of ENC (Fig.1), a similar dependence of the final states of particle 2 on the magnitude of the magnetic field B will be discovered.

I would like to thank Kenneth Augustyn for bringing the problem of ENC to my attention.

\begin{flushleft}
	
\end{flushleft}

\end{document}